\documentclass[conference]{IEEEtran}
\IEEEoverridecommandlockouts

\usepackage{natbib}
\usepackage{amsmath,amssymb,amsfonts}
\usepackage{algorithmic}
\usepackage{graphicx}
\usepackage{textcomp}
\usepackage{caption}
\usepackage{subcaption}
\usepackage{tabularx}
\usepackage{array}
\usepackage{booktabs}
\usepackage{multirow}
\usepackage{float}
\usepackage{url}
\usepackage{makecell,siunitx}
\usepackage[table]{xcolor}
\usepackage[hidelinks]{hyperref}
\usepackage{fancyhdr}

\fancypagestyle{obpdcfooter}{
    \fancyhf{}
    \fancyfoot[C]{\footnotesize \textbf{OBPDC 2026}}

}

\renewcommand{\arraystretch}{1.2}
\newcolumntype{Y}{>{\centering\arraybackslash}X}
\newcolumntype{L}{>{\raggedright\arraybackslash}X}
\DeclareSIUnit\px{px}
\newcommand{\degradimg}[1]{%
    \raisebox{0pt}[\dimexpr\height+4pt\relax][\dimexpr\depth+4pt\relax]{%
        \includegraphics[width=0.195\textwidth]{#1}%
    }%
}

\newcommand{\figSimExamples}{%
\begin{figure*}[t]
    \centering

    \setlength{\tabcolsep}{3pt}
    \renewcommand{\arraystretch}{1.15}
    \setlength{\arrayrulewidth}{0.5pt}

    \begin{tabular}{|c|c|c|c|c|}
        \hline

        \textbf{GSD} &
        \textbf{GSD only} &
        \textbf{MTF only} &
        \textbf{SNR only} &
        \textbf{MTF + SNR}
        \\

        \cline{2-5}

        &
        -- &
        $\mathrm{MTF}_{\mathrm{Nyq}}=0.03$ &
        $\mathrm{SNR}=(20,50)$ &
        \shortstack{
            $\mathrm{MTF}_{\mathrm{Nyq}}=0.03$\\
            $\mathrm{SNR}=(20,50)$
        }
        \\

        \hline

        \rotatebox{90}{\textbf{GSD 100 cm}}
        &
        \degradimg{
            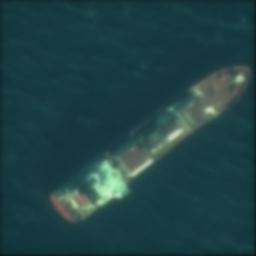
        }
        &
        \degradimg{
            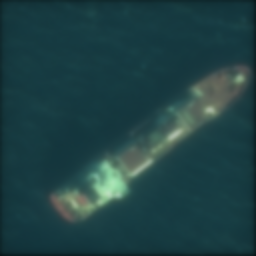
        }
        &
        \degradimg{
            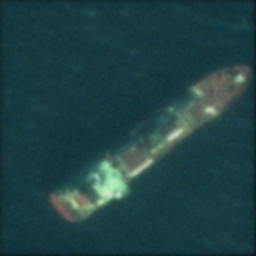
        }
        &
        \degradimg{
            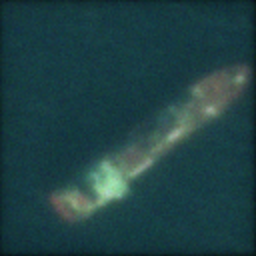
        }
        \\

        \hline

        \rotatebox{90}{\textbf{GSD 200 cm}}
        &
        \degradimg{
            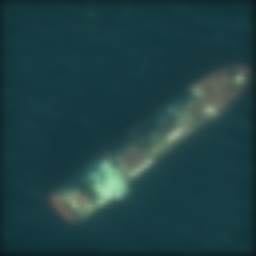
        }
        &
        \degradimg{
            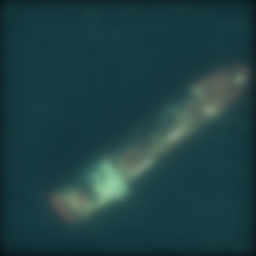
        }
        &
        \degradimg{
            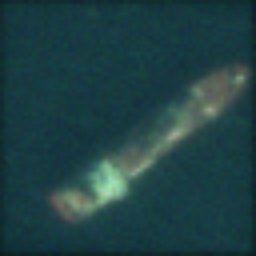
        }
        &
        \degradimg{
            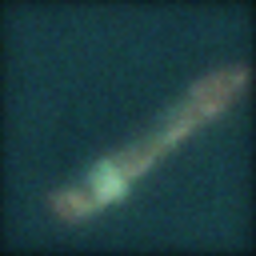
        }
        \\

        \hline
    \end{tabular}

    \caption{Examples of GSD, MTF, SNR, and combined degradations at GSDs of 100 and 200~cm.}
    \label{fig:degradation_examples}

\end{figure*}%
}

\newcommand{\tabResults}{%

\begin{table*}[t]
    \centering
    \caption{Detection performance at GSD = 100~cm.}
    \label{tab:all_results_gsd100}
    \scriptsize
    \setlength{\tabcolsep}{2.25pt}
    \renewcommand{\arraystretch}{1.03}

    \resizebox{\textwidth}{!}{%
    \begin{tabular}{llcc ccc ccc ccc}
        \toprule

        \multirow{2}{*}{Degradation} &
        \multirow{2}{*}{Experiment} &
        \multirow{2}{*}{$\mathrm{MTF}_{\mathrm{Nyq}}$} &
        \multirow{2}{*}{SNR $(L_0,L_1)$ [dB]} &
        \multicolumn{3}{c}{NanoDet \cite{nanodet}} &
        \multicolumn{3}{c}{YOLOv5s \cite{yolov5}} &
        \multicolumn{3}{c}{YOLOX-S \cite{yolox}} \\

        \cmidrule(lr){5-7}
        \cmidrule(lr){8-10}
        \cmidrule(lr){11-13}

        & & & &
        AP & AP$_{50}$ & AP$_{75}$ &
        AP & AP$_{50}$ & AP$_{75}$ &
        AP & AP$_{50}$ & AP$_{75}$ \\

        \midrule

        MTF only & mtf\_025\_gsd100 & 0.250 & (100, 250)
        & 0.244 & 0.305 & 0.267
        & 0.345 & 0.412 & 0.383
        & 0.281 & 0.342 & 0.318 \\

        MTF only & mtf\_015\_gsd100 & 0.150 & (100, 250)
        & 0.222 & 0.276 & 0.241
        & 0.298 & 0.370 & 0.331
        & 0.278 & 0.338 & 0.312 \\

        MTF only & mtf\_010\_gsd100 & 0.100 & (100, 250)
        & 0.221 & 0.271 & 0.239
        & 0.317 & 0.383 & 0.352
        & 0.263 & 0.336 & 0.298 \\

        MTF only & mtf\_007\_gsd100 & 0.070 & (100, 250)
        & 0.218 & 0.265 & 0.235
        & 0.298 & 0.364 & 0.331
        & 0.268 & 0.333 & 0.303 \\

        MTF only & mtf\_003\_gsd100 & 0.030 & (100, 250)
        & 0.215 & 0.264 & 0.231
        & 0.272 & 0.336 & 0.302
        & 0.238 & 0.297 & 0.270 \\

        MTF only & mtf\_0015\_gsd100 & 0.015 & (100, 250)
        & 0.198 & 0.244 & 0.213
        & 0.265 & 0.332 & 0.294
        & 0.241 & 0.317 & 0.277 \\

        MTF only & mtf\_0010\_gsd100 & 0.010 & (100, 250)
        & 0.192 & 0.241 & 0.209
        & 0.290 & 0.348 & 0.315
        & 0.253 & 0.323 & 0.284 \\

        MTF only & mtf\_0005\_gsd100 & 0.005 & (100, 250)
        & 0.191 & 0.246 & 0.212
        & 0.258 & 0.327 & 0.291
        & 0.228 & 0.296 & 0.254 \\

        \midrule

        SNR only & snr\_080\_gsd100 & 0.100 & (80, 170)
        & 0.215 & 0.263 & 0.232
        & 0.302 & 0.370 & 0.331
        & 0.260 & 0.337 & 0.282 \\

        SNR only & snr\_072\_gsd100 & 0.100 & (72, 175)
        & 0.245 & 0.295 & 0.266
        & 0.286 & 0.350 & 0.318
        & 0.236 & 0.302 & 0.264 \\

        SNR only & snr\_050\_gsd100 & 0.100 & (50, 110)
        & 0.205 & 0.258 & 0.223
        & 0.296 & 0.356 & 0.326
        & 0.265 & 0.343 & 0.302 \\

        SNR only & snr\_037\_gsd100 & 0.100 & (37, 85)
        & 0.240 & 0.296 & 0.260
        & 0.296 & 0.367 & 0.329
        & 0.265 & 0.326 & 0.300 \\

        SNR only & snr\_026\_gsd100 & 0.100 & (26, 60)
        & 0.221 & 0.271 & 0.238
        & 0.285 & 0.355 & 0.317
        & 0.268 & 0.340 & 0.306 \\

        SNR only & snr\_020\_gsd100 & 0.100 & (20, 50)
        & 0.230 & 0.284 & 0.250
        & 0.304 & 0.372 & 0.329
        & 0.278 & 0.348 & 0.315 \\

        SNR only & snr\_014\_gsd100 & 0.100 & (14, 29)
        & 0.215 & 0.266 & 0.233
        & 0.269 & 0.337 & 0.299
        & 0.258 & 0.323 & 0.292 \\

        SNR only & snr\_010\_gsd100 & 0.100 & (10, 20)
        & 0.219 & 0.278 & 0.240
        & 0.266 & 0.332 & 0.295
        & 0.257 & 0.336 & 0.292 \\

        SNR only & snr\_007\_gsd100 & 0.100 & (7, 14)
        & 0.215 & 0.273 & 0.234
        & 0.252 & 0.313 & 0.279
        & 0.230 & 0.306 & 0.262 \\

        SNR only & snr\_005\_gsd100 & 0.100 & (5, 10)
        & 0.185 & 0.239 & 0.200
        & 0.259 & 0.331 & 0.293
        & 0.234 & 0.315 & 0.274 \\

        \midrule

        MTF + SNR & mix\_025\_080\_gsd100 & 0.250 & (80, 170)
        & 0.218 & 0.274 & 0.241
        & 0.330 & 0.401 & 0.363
        & 0.240 & 0.298 & 0.270 \\

        MTF + SNR & mix\_015\_072\_gsd100 & 0.150 & (72, 175)
        & 0.237 & 0.287 & 0.253
        & 0.306 & 0.380 & 0.342
        & 0.247 & 0.307 & 0.279 \\

        MTF + SNR & mix\_010\_037\_gsd100 & 0.100 & (37, 85)
        & 0.230 & 0.281 & 0.248
        & 0.316 & 0.379 & 0.349
        & 0.243 & 0.299 & 0.272 \\

        MTF + SNR & mix\_007\_026\_gsd100 & 0.070 & (26, 60)
        & 0.225 & 0.282 & 0.245
        & 0.299 & 0.368 & 0.331
        & 0.231 & 0.290 & 0.259 \\

        MTF + SNR & mix\_003\_low\_gsd100 & 0.030 & (20, 50)
        & 0.186 & 0.235 & 0.205
        & 0.281 & 0.343 & 0.306
        & 0.238 & 0.305 & 0.267 \\

        MTF + SNR & mix\_0015\_014\_gsd100 & 0.015 & (14, 29)
        & 0.217 & 0.270 & 0.237
        & 0.230 & 0.300 & 0.261
        & 0.242 & 0.315 & 0.273 \\

        MTF + SNR & mix\_0010\_007\_gsd100 & 0.010 & (7, 14)
        & 0.178 & 0.233 & 0.198
        & 0.209 & 0.266 & 0.233
        & 0.201 & 0.270 & 0.229 \\

        MTF + SNR & mix\_0005\_005\_gsd100 & 0.005 & (5, 10)
        & 0.150 & 0.201 & 0.164
        & 0.170 & 0.230 & 0.189
        & 0.189 & 0.251 & 0.218 \\

        \bottomrule
    \end{tabular}%
    }

\end{table*}

\begin{table*}[t]
    \centering
    \caption{Detection performance at GSD = 200~cm.}
    \label{tab:all_results_gsd200}
    \scriptsize
    \setlength{\tabcolsep}{2.25pt}
    \renewcommand{\arraystretch}{1.03}

    \resizebox{\textwidth}{!}{%
    \begin{tabular}{llcc ccc ccc ccc}
        \toprule

        \multirow{2}{*}{Degradation} &
        \multirow{2}{*}{Experiment} &
        \multirow{2}{*}{$\mathrm{MTF}_{\mathrm{Nyq}}$} &
        \multirow{2}{*}{SNR $(L_0,L_1)$ [dB]} &
        \multicolumn{3}{c}{NanoDet \cite{nanodet}} &
        \multicolumn{3}{c}{YOLOv5s \cite{yolov5}} &
        \multicolumn{3}{c}{YOLOX-S \cite{yolox}} \\

        \cmidrule(lr){5-7}
        \cmidrule(lr){8-10}
        \cmidrule(lr){11-13}

        & & & &
        AP & AP$_{50}$ & AP$_{75}$ &
        AP & AP$_{50}$ & AP$_{75}$ &
        AP & AP$_{50}$ & AP$_{75}$ \\

        \midrule

        MTF only & mtf\_025\_gsd200 & 0.250 & (100, 250)
        & 0.176 & 0.220 & 0.190
        & 0.248 & 0.311 & 0.274
        & 0.210 & 0.267 & 0.239 \\

        MTF only & mtf\_015\_gsd200 & 0.150 & (100, 250)
        & 0.166 & 0.206 & 0.178
        & 0.246 & 0.298 & 0.267
        & 0.211 & 0.278 & 0.241 \\

        MTF only & mtf\_010\_gsd200 & 0.100 & (100, 250)
        & 0.198 & 0.246 & 0.216
        & 0.256 & 0.320 & 0.282
        & 0.236 & 0.301 & 0.266 \\

        MTF only & mtf\_007\_gsd200 & 0.070 & (100, 250)
        & 0.179 & 0.229 & 0.196
        & 0.242 & 0.299 & 0.260
        & 0.229 & 0.296 & 0.260 \\

        MTF only & mtf\_003\_gsd200 & 0.030 & (100, 250)
        & 0.184 & 0.231 & 0.196
        & 0.245 & 0.302 & 0.268
        & 0.209 & 0.279 & 0.239 \\

        MTF only & mtf\_0015\_gsd200 & 0.015 & (100, 250)
        & 0.177 & 0.225 & 0.193
        & 0.227 & 0.278 & 0.246
        & 0.211 & 0.273 & 0.237 \\

        MTF only & mtf\_0010\_gsd200 & 0.010 & (100, 250)
        & 0.165 & 0.208 & 0.176
        & 0.235 & 0.301 & 0.266
        & 0.185 & 0.251 & 0.210 \\

        MTF only & mtf\_0005\_gsd200 & 0.005 & (100, 250)
        & 0.180 & 0.229 & 0.196
        & 0.198 & 0.256 & 0.226
        & 0.216 & 0.286 & 0.244 \\

        \midrule

        SNR only & snr\_080\_gsd200 & 0.100 & (80, 170)
        & 0.205 & 0.2544 & 0.225
        & 0.246 & 0.318 & 0.281
        & 0.203 & 0.269 & 0.231 \\

        SNR only & snr\_072\_gsd200 & 0.100 & (72, 175)
        & 0.198 & 0.246 & 0.217
        & 0.242 & 0.301 & 0.263
        & 0.196 & 0.253 & 0.225 \\

        SNR only & snr\_050\_gsd200 & 0.100 & (50, 110)
        & 0.205 & 0.258 & 0.224
        & 0.265 & 0.333 & 0.292
        & 0.218 & 0.281 & 0.251 \\

        SNR only & snr\_037\_gsd200 & 0.100 & (37, 85)
        & 0.199 & 0.249 & 0.214
        & 0.239 & 0.306 & 0.263
        & 0.196 & 0.256 & 0.218 \\

        SNR only & snr\_026\_gsd200 & 0.100 & (26, 60)
        & 0.180 & 0.230 & 0.200
        & 0.231 & 0.302 & 0.261
        & 0.231 & 0.302 & 0.263 \\

        SNR only & snr\_020\_gsd200 & 0.100 & (20, 50)
        & 0.202 & 0.250 & 0.219
        & 0.238 & 0.297 & 0.265
        & 0.213 & 0.282 & 0.234 \\

        SNR only & snr\_014\_gsd200 & 0.100 & (14, 29)
        & 0.167 & 0.211 & 0.178
        & 0.220 & 0.290 & 0.253
        & 0.174 & 0.234 & 0.200 \\

        SNR only & snr\_010\_gsd200 & 0.100 & (10, 20)
        & 0.179 & 0.229 & 0.194
        & 0.217 & 0.277 & 0.242
        & 0.205 & 0.273 & 0.230 \\

        SNR only & snr\_007\_gsd200 & 0.100 & (7, 14)
        & 0.176 & 0.227 & 0.193
        & 0.166 & 0.223 & 0.190
        & 0.179 & 0.252 & 0.202 \\

        SNR only & snr\_005\_gsd200 & 0.100 & (5, 10)
        & 0.153 & 0.193 & 0.163
        & 0.164 & 0.227 & 0.191
        & 0.181 & 0.241 & 0.206 \\

        \midrule

        MTF + SNR & mix\_025\_080\_gsd200 & 0.250 & (80, 170)
        & 0.203 & 0.251 & 0.217
        & 0.244 & 0.297 & 0.268
        & 0.219 & 0.278 & 0.246 \\

        MTF + SNR & mix\_015\_072\_gsd200 & 0.150 & (72, 175)
        & 0.191 & 0.237 & 0.209
        & 0.242 & 0.302 & 0.266
        & 0.219 & 0.276 & 0.245 \\

        MTF + SNR & mix\_010\_037\_gsd200 & 0.100 & (37, 85)
        & 0.184 & 0.232 & 0.199
        & 0.235 & 0.304 & 0.263
        & 0.204 & 0.266 & 0.232 \\

        MTF + SNR & mix\_007\_026\_gsd200 & 0.070 & (26, 60)
        & 0.175 & 0.226 & 0.193
        & 0.229 & 0.290 & 0.247
        & 0.224 & 0.287 & 0.249 \\

        MTF + SNR & mix\_003\_low\_gsd200 & 0.030 & (20, 50)
        & 0.185 & 0.245 & 0.209
        & 0.205 & 0.264 & 0.228
        & 0.206 & 0.273 & 0.236 \\

        MTF + SNR & mix\_0015\_014\_gsd200 & 0.015 & (14, 29)
        & 0.156 & 0.200 & 0.166
        & 0.186 & 0.246 & 0.210
        & 0.185 & 0.244 & 0.208 \\

        MTF + SNR & mix\_0010\_007\_gsd200 & 0.010 & (7, 14)
        & 0.091 & 0.127 & 0.096
        & 0.124 & 0.170 & 0.138
        & 0.151 & 0.204 & 0.172 \\

        MTF + SNR & mix\_0005\_005\_gsd200 & 0.005 & (5, 10)
        & 0.097 & 0.131 & 0.104
        & 0.120 & 0.166 & 0.134
        & 0.115 & 0.168 & 0.130 \\

        \bottomrule
    \end{tabular}%
    }

\end{table*}
}

\begin{document}
\pagestyle{obpdcfooter}

\title{Raw Imagery Impacting Your AI: Should You Care?}

\author{
    \IEEEauthorblockN{
        Adrien Dorise \IEEEauthorrefmark{1}\IEEEauthorrefmark{2},
        Marjorie Bellizzi \IEEEauthorrefmark{2},
        St\'ephane May \IEEEauthorrefmark{1}
    }

    \IEEEauthorblockA{\IEEEauthorrefmark{1}CNES, Toulouse, France\\
    Emails: \{adrien.dorise, stephane.may\}@cnes.fr}

    \IEEEauthorblockA{\IEEEauthorrefmark{2}IRT Saint-Exup\'ery, Toulouse, France\\
    Email: marjorie.bellizzi@irt-saintexupery.com}
}

\maketitle
\thispagestyle{obpdcfooter}

\begin{abstract}
Onboard AI is gaining interest for space applications such as vessel, wildfire, and cloud detection, where real-time processing can improve mission reactivity and reduce downlink needs. However, onboard models may operate on raw or minimally processed imagery rather than on restored ground products. This study evaluates how image degradation affects object detection by varying Signal-to-Noise Ratio (SNR), Modulation Transfer Function (MTF) at Nyquist, and Ground Sampling Distance (GSD). Controlled degradations are applied to Very High Resolution Maxar imagery, and three lightweight detectors, YOLOv5s, YOLOX-S, and NanoDet, are evaluated on the resulting operating points. The results show that the impact of image quality depends on the degradation mechanism, and that increasing degradation does not necessarily lead to a proportional decrease in vessel detection performance. GSD produces the most consistent performance shift, while MTF and SNR effects depend more on the model and resolution. Severe combinations of blur and noise produce the largest losses. These results provide task-level information that can support sensor, processing, and AI trade-offs for future onboard systems.

\end{abstract}


\section{Introduction}

The computational cost of modern artificial intelligence (AI) algorithms has historically limited their use in space applications, leaving most image interpretation to ground processing. This situation is changing as increasingly capable processors are evaluated for space use and recent missions demonstrate onboard inference capabilities \cite{irma_imagini,opssat_meoni,cloud_detection_onboard, Goudemant_2026_CVPR}. Running AI directly onboard can improve mission reactivity, reduce downlink volume, and prioritise communication and storage resources for the most useful observations \cite{remote_sensing_review,pyraws_thraws}. These capabilities are particularly attractive for time-sensitive Earth-observation applications such as vessel, wildfire, and cloud detection.

Most remote-sensing AI pipelines nevertheless assume images that have already undergone a conventional ground-processing chain. Level-1 products can include radiometric and geometric corrections, coregistration, resampling, and image restoration \cite{pleiades_restoration,venus_dataset_full_author}. Reproducing the complete processing chain onboard may be undesirable when power, memory, latency, and computing resources are constrained \cite{remote_sensing_review,jetson_space,versal_spaice_project, CIAR}. An alternative is to perform inference earlier in the processing chain using imagery closer to the sensor output, an approach increasingly investigated for onboard Earth-observation applications \cite{pyraws_thraws,pyraws_maritime1,venus_dataset_full_author}.

This introduces a trade-off between image quality and onboard processing complexity. Processing stages that are important for producing high-quality images for human interpretation do not necessarily provide the same benefit to an AI model. Conversely, image degradations that appear visually moderate may remove spatial or radiometric information important for detection. Consequently, sensor and processing requirements derived exclusively from conventional image-quality criteria may not directly represent the requirements of the downstream AI task.

This raises a practical system-design question: \emph{how much image quality is actually required by the downstream AI task?} Raw optical imagery may suffer from reduced spatial sampling, loss of spatial contrast, and radiometric noise. These effects can be represented through the Ground Sampling Distance (GSD), Modulation Transfer Function (MTF), and Signal-to-Noise Ratio (SNR) \cite{pleiades_restoration,raw_detection_edhpc}. However, their effect on a learned detector is not necessarily proportional to their effect on visual image quality. A detector may tolerate a range of degradation before losing accuracy rapidly once task-relevant information becomes insufficient. Identifying such regimes can help determine whether sensor specifications or onboard restoration requirements can be relaxed without significantly affecting the mission-level AI task.

In previous work, we developed a pipeline that generates simulated raw-like products from high-resolution Level-1 imagery \cite{raw_detection_edhpc} and studied lightweight onboard image restoration \cite{convbeers}. This study focuses on task-level sensitivity to controlled variations in GSD, MTF, and SNR. Experiments are performed at GSDs of 100 and 200~cm. MTF-only and SNR-only sweeps are separated from a combined MTF and SNR sweep to distinguish the effects of each degradation family. Three lightweight detectors, NanoDet, YOLOv5s, and YOLOX-S, are compared to evaluate architecture-dependent responses. The objective is not to define a universal optical-quality threshold, but to provide task-level evidence that can support sensor, processor, and AI co-design.

\section{Materials and Methods}

\subsection{Image degradation simulation}
\label{sec:simulation}

To evaluate object detection under controlled image-quality degradations, we generate observations from reference images using a sensor simulation pipeline. No restoration is considered after degradation. The simulation consists of three successive operations: MTF degradation, GSD modification, and signal-dependent noise injection, following the degradation model introduced in our previous work \cite{raw_detection_edhpc}

\subsubsection{Spatial degradation through the MTF}

A parametric two-dimensional modulation transfer function represents optical-system and detector-induced loss of spatial contrast. In the frequency domain, the MTF is defined as

\begin{equation}
\mathrm{MTF}(f_x,f_y)
=
\exp(-\gamma f_r)
\,\mathrm{sinc}(f_x)
\,\mathrm{sinc}(f_y),
\label{eq:mtf}
\end{equation}

with

\begin{equation}
f_r = \sqrt{f_x^2 + f_y^2},
\end{equation}

where the sinc terms approximate the detector response along the two spatial axes, while the exponential term represents the optical contribution. The parameter $\gamma$ is determined from a prescribed MTF value at the Nyquist frequency, $\mathrm{MTF}_{\mathrm{Nyq}}$, according to

\begin{equation}
\gamma
=
-2
\log
\left(
\frac{\mathrm{MTF}_{\mathrm{Nyq}}}
{\mathrm{sinc}(0.5)}
\right).
\label{eq:gamma}
\end{equation}

The MTF is sampled over an oversampled frequency grid and converted into its corresponding point-spread function (PSF) through an inverse Fourier transform,

\begin{equation}
\mathrm{PSF}
=
\mathcal{F}^{-1}
\left\{
\mathrm{MTF}
\right\}.
\end{equation}

The resulting PSF is convolved independently with each spectral channel of the input image using FFT-based convolution.

\subsubsection{GSD degradation}

After applying MTF degradation, we modify the spatial sampling to reproduce the target sensor's Ground Sampling Distance. Let $\mathrm{GSD}_{\mathrm{src}}$ and $\mathrm{GSD}_{\mathrm{target}}$ denote the spatial resolutions of the reference and simulated images. The downsampling ratio is defined as

\begin{equation}
r
=
\frac{\mathrm{GSD}_{\mathrm{target}}}
{\mathrm{GSD}_{\mathrm{src}}}.
\label{eq:gsd_ratio}
\end{equation}

First, the image dimensions are adjusted to match the resampling ratio, after which the image is spatially subsampled with a centred offset to avoid an additional spatial shift. When geospatial metadata are available, the image dimensions and affine transform are updated consistently with the new GSD.

Separating the MTF and GSD transformations lets us model two spatial effects explicitly. The MTF represents attenuation of spatial frequencies caused by the imaging system, while the GSD update represents discrete sampling of the ground scene.

\subsubsection{Signal-dependent radiometric noise}

We simulate radiometric degradation using a signal-dependent Gaussian noise model. The noise variance is assumed to depend linearly on the signal level $L$:

\begin{equation}
\sigma^2(L)
=
\alpha L + \beta,
\label{eq:noise_model}
\end{equation}

where $\alpha$ and $\beta$ control the dependence of the noise variance on signal intensity. These parameters are determined from two reference operating points, $(L_0,\mathrm{SNR}_0)$ and $(L_1,\mathrm{SNR}_1)$. The corresponding noise variances are

\begin{equation}
\sigma_0^2
=
\left(
\frac{L_0}{\mathrm{SNR}_0}
\right)^2,
\end{equation}

and

\begin{equation}
\sigma_1^2
=
\left(
\frac{L_1}{\mathrm{SNR}_1}
\right)^2.
\end{equation}

The coefficients of the linear noise model are then

\begin{equation}
\alpha
=
\frac{\sigma_0^2-\sigma_1^2}
{L_0-L_1},
\label{eq:alpha}
\end{equation}

and

\begin{equation}
\beta
=
\sigma_0^2-\alpha L_0.
\label{eq:beta}
\end{equation}

For each pixel, the noise standard deviation is evaluated as

\begin{equation}
\sigma(L)
=
\sqrt{
\max
\left(
0,\alpha L+\beta
\right)
},
\label{eq:sigma}
\end{equation}

and an independent Gaussian noise sample is sampled according to

\begin{equation}
n
\sim
\mathcal{N}
\left(
0,\sigma^2(L)
\right).
\end{equation}

The simulated noisy image is therefore

\begin{equation}
I_{\mathrm{noisy}}
=
I_{\mathrm{GSD}} + n.
\end{equation}

This formulation produces a noise level that varies with the image signal rather than assuming constant variance over the complete image. The resulting values are finally clipped to the sensor digital range.

The complete degradation pipeline can be summarized as

\begin{equation}
I_{\mathrm{ref}}
\xrightarrow{\mathrm{MTF}}
I_{\mathrm{MTF}}
\xrightarrow{\mathrm{GSD}}
I_{\mathrm{GSD}}
\xrightarrow{\mathrm{SNR}}
I_{\mathrm{sim}},
\label{eq:simulation_pipeline}
\end{equation}

where $I_{\mathrm{ref}}$ denotes the reference image and $I_{\mathrm{sim}}$ the final simulated sensor observation.

\subsection{Dataset}

The experiments rely on a custom high-resolution database designed for the development and evaluation of AI-based ship-detection algorithms. The dataset contains 47 high-resolution scenes with an original GSD of 50~cm and more than 24,000 annotated ships distributed over 53 vessel classes, including small boats, ore carriers, and commercial cargo ships. The imagery is derived from Maxar Standard 2A products and includes one panchromatic band together with red, green, and blue multispectral bands acquired at a coarser native resolution. We pansharpen the data using the weighted Brovey method \cite{pleiades_restoration} to generate the final 50~cm images. The products also include standard radiometric and geometric corrections.

\subsection{Detection models}

Three lightweight object-detection architectures are considered: YOLOv5s \cite{yolov5}, NanoDet \cite{nanodet}, and YOLOX-S \cite{yolox}. We selected these models because onboard processing constrains inference latency, memory footprint, and deployment complexity \cite{remote_sensing_review,jetson_space,versal_spaice_project}. Lightweight object detection has also been investigated specifically for onboard vessel-detection scenarios \cite{versal_ship_detection_yolo,CIAR}.

We train all models for 100 epochs using two NVIDIA RTX~4090 GPU, with an average training time of approximately 10 hours per model and operating point. Detection performance is reported using COCO-style average precision (AP), AP$_{50}$, and AP$_{75}$.

\subsection{Experimental design}
\label{sec:experiments}

\figSimExamples

The original imagery, with a reference GSD of approximately 50~cm, is retained as the high-resolution source condition but is not included in the controlled MTF and SNR grid. In the current simulation pipeline, independent control of the prescribed MTF and SNR values is associated with generation at a new target GSD. When we retain the original GSD, the effective MTF and SNR remain inherited from the source imagery.

We therefore perform controlled experiments at target GSDs of 100 and 200~cm. For each GSD, we first vary MTF and SNR independently to isolate spatial blur and radiometric noise. A third sweep jointly varies MTF and SNR to reproduce combined degradation. The complete set of operating points is reported in Table~\ref{tab:all_results_gsd100} and Table~\ref{tab:all_results_gsd200}.

The nominal simulated configuration uses $\mathrm{MTF}_{\mathrm{Nyq}}=0.10$ and SNR parameters $(100,250)$. The MTF-only sweep varies $\mathrm{MTF}_{\mathrm{Nyq}}$ from 0.25 to 0.005 while keeping the nominal SNR. The SNR-only sweep progressively reduces the two SNR parameters while retaining $\mathrm{MTF}_{\mathrm{Nyq}}=0.10$. Finally, eight combined operating points vary both quantities simultaneously. We deliberately chose severe degradation parameters to visualise the impact of such degradation on detection models clearly.  Representative image crops are shown in Fig.~\ref{fig:degradation_examples}.

%\tabSimParam

\section{Results}

Tables~\ref{tab:all_results_gsd100} and~\ref{tab:all_results_gsd200} report AP, AP$_{50}$, and AP$_{75}$ for all available configurations. The following analysis focuses mainly on AP$_{50}$ and on the overall trends across the degradation sweeps.

\subsection{Effect of ground sampling distance}

\begin{figure}[htbp]
    \centering
    \includegraphics[width=0.48\textwidth]{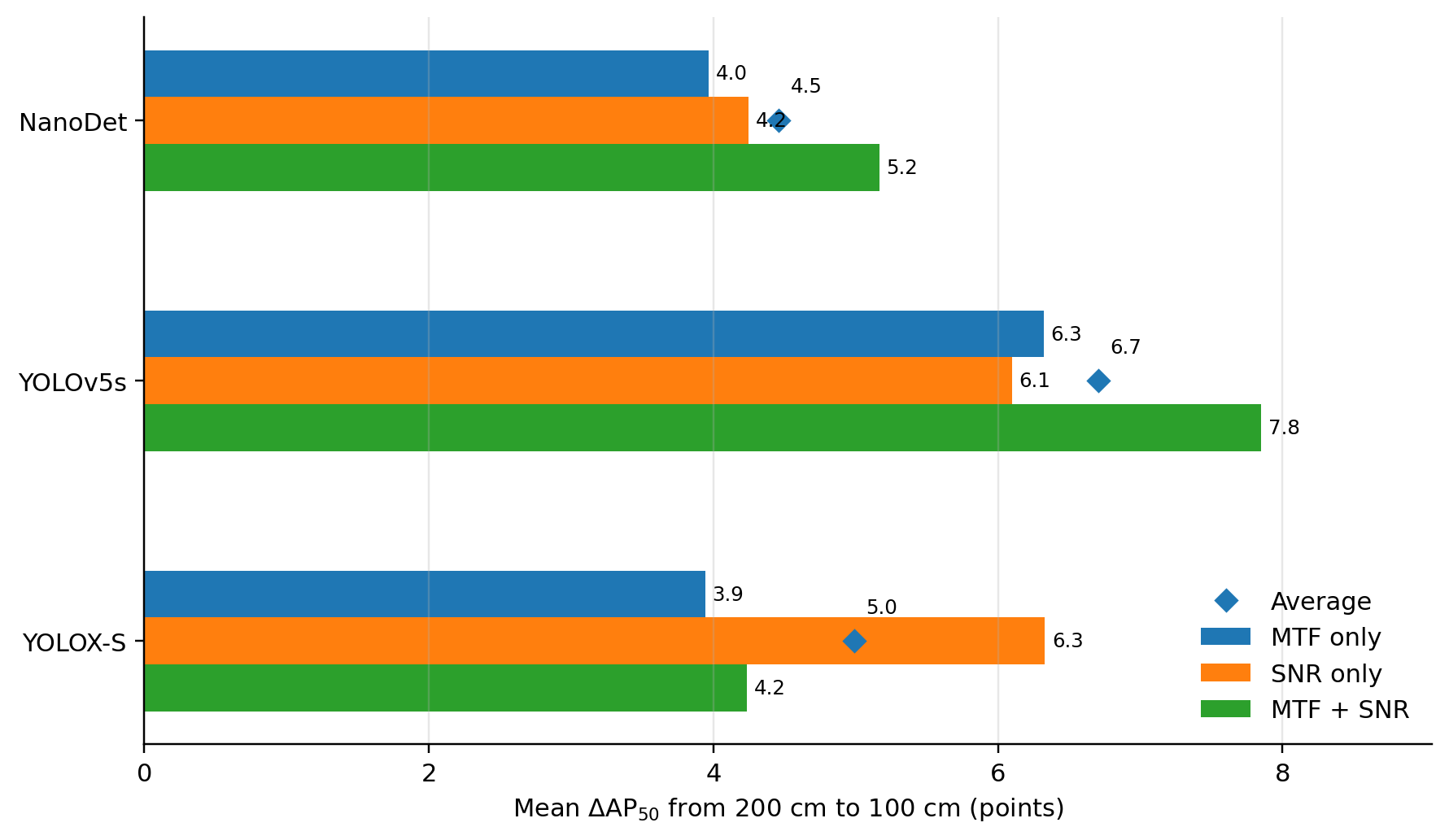} 
    \caption{Mean AP$_{50}$ gain from GSD 200cm to GSD 100cm (in percentage points)} 
    \label{fig:gsd_gain} 
\end{figure}

As shown in Fig.~\ref{fig:gsd_gain}, increasing the GSD from 100 to 200~cm consistently reduces detection performance, with losses ranging from 3.9 to 7.8 percentage points depending on the degradation configuration. For YOLOv5s and YOLOX-S, AP$_{50}$ is lower at 200~cm for all configurations available at both resolutions. NanoDet follows the same overall trend, with only one small reversal in the combined MTF and SNR sweep.

The effect is visible across MTF-only, SNR-only, and combined degradation conditions, indicating that spatial sampling remains an important factor independently of the other image-quality parameters. However, the magnitude of the loss depends on both the detector and the operating point.

\subsection{MTF-only degradation}

At GSD~=~100~cm, lower MTF generally reduces detection performance. This trend is especially clear for YOLOv5s and NanoDet, where the lowest MTF configurations perform noticeably worse than the highest-quality configurations. YOLOX-S shows the same general tendency over its available operating points, although the evolution is not strictly monotonic.

At GSD~=~200~cm, the relationship between MTF and detection performance becomes less regular. Very low MTF values tend to reduce performance, but neighbouring operating points frequently change order. These results show that MTF degradation is harmful at sufficiently low values, while moderate changes in MTF do not translate directly into proportional changes in detection accuracy.

\subsection{SNR-only degradation}

At GSD~=~100~cm, detection performance remains relatively stable over much of the tested SNR range. The lowest SNR configurations generally lead to poorer results, but variations between intermediate operating points are small and non-monotonic for all three detectors. This suggests that, at this GSD, moderate radiometric degradation is comparatively well tolerated.

The effect of SNR becomes more visible at GSD~=~200~cm. YOLOv5s shows a clear reduction in AP$_{50}$ toward the lowest SNR values, while NanoDet also performs worse under severe noise. YOLOX-S shows more variation between intermediate configurations, but its lowest-SNR results remain below its best SNR-only operating points. Overall, the results suggest that sensitivity to radiometric noise is mostly stable for low to middle SNR degradations, but heavily noisy images suffer greater performance loss.

\subsection{Combined MTF and SNR degradation}

The strongest performance losses are observed when severe MTF and SNR degradation occur simultaneously. For all three detectors, the lowest-quality combined configurations produce substantially lower AP$_{50}$ than the corresponding MTF-only or SNR-only cases.

This effect is particularly visible at GSD~=~200~cm, where the final two combined operating points show a sharp performance drop across all three architectures. A similar trend is observed at 100~cm, although the intermediate configurations remain less regular.

The combined results therefore identify severe blur and noise as the most challenging operating regime considered in this study. The response is not perfectly monotonic, especially at intermediate quality levels, but performance clearly deteriorates once both degradation mechanisms become strong.

\begin{figure*}[t]
    \centering

    \begin{subfigure}[t]{0.49\textwidth}
        \centering
        \includegraphics[height=0.19\textheight,keepaspectratio]{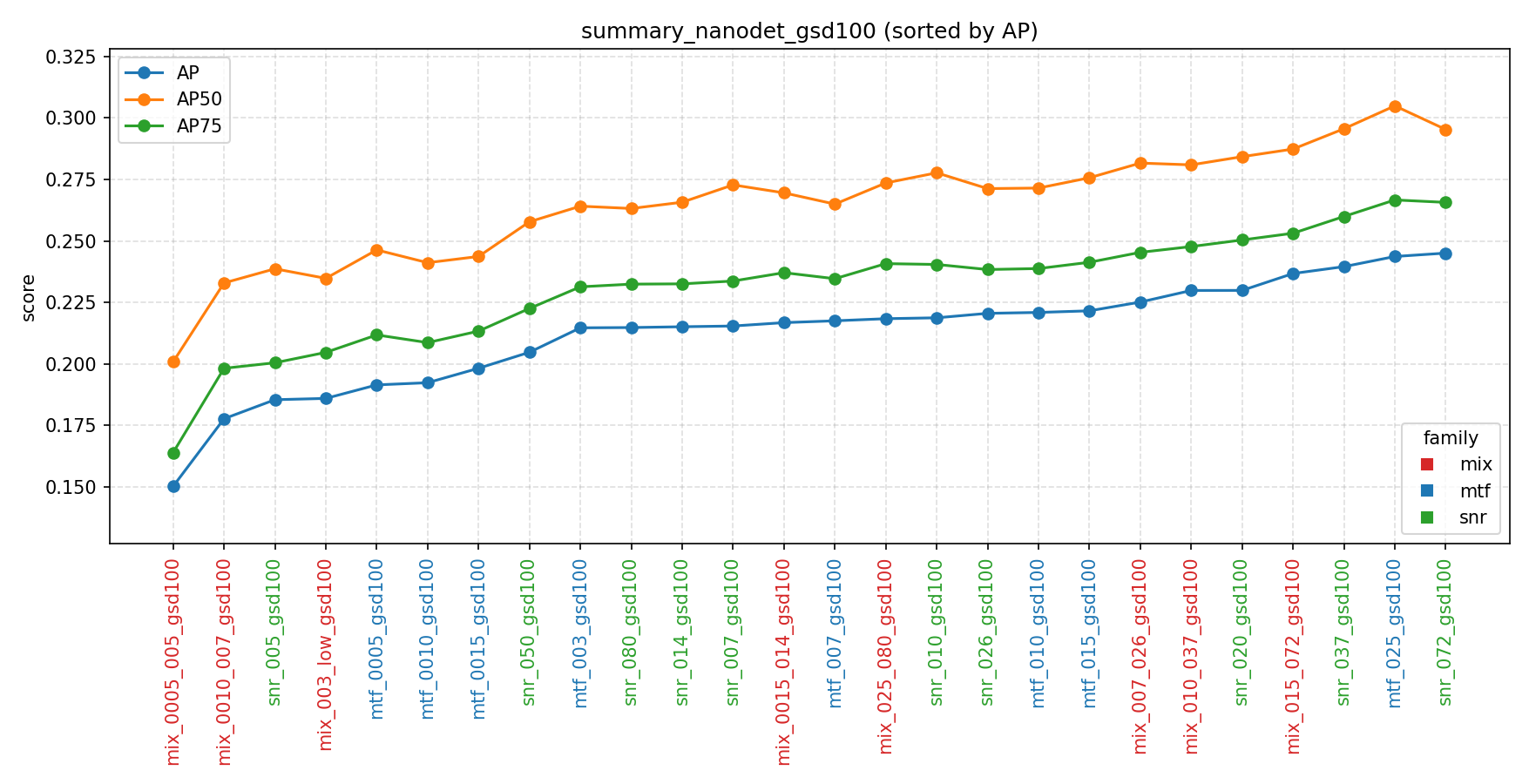}
        \caption{NanoDet configurations, ordered by AP.}
        \label{fig:nanodet_gsd100_sorted}
    \end{subfigure}
    \hfill
    \begin{subfigure}[t]{0.49\textwidth}
        \centering
        \includegraphics[height=0.19\textheight,keepaspectratio]{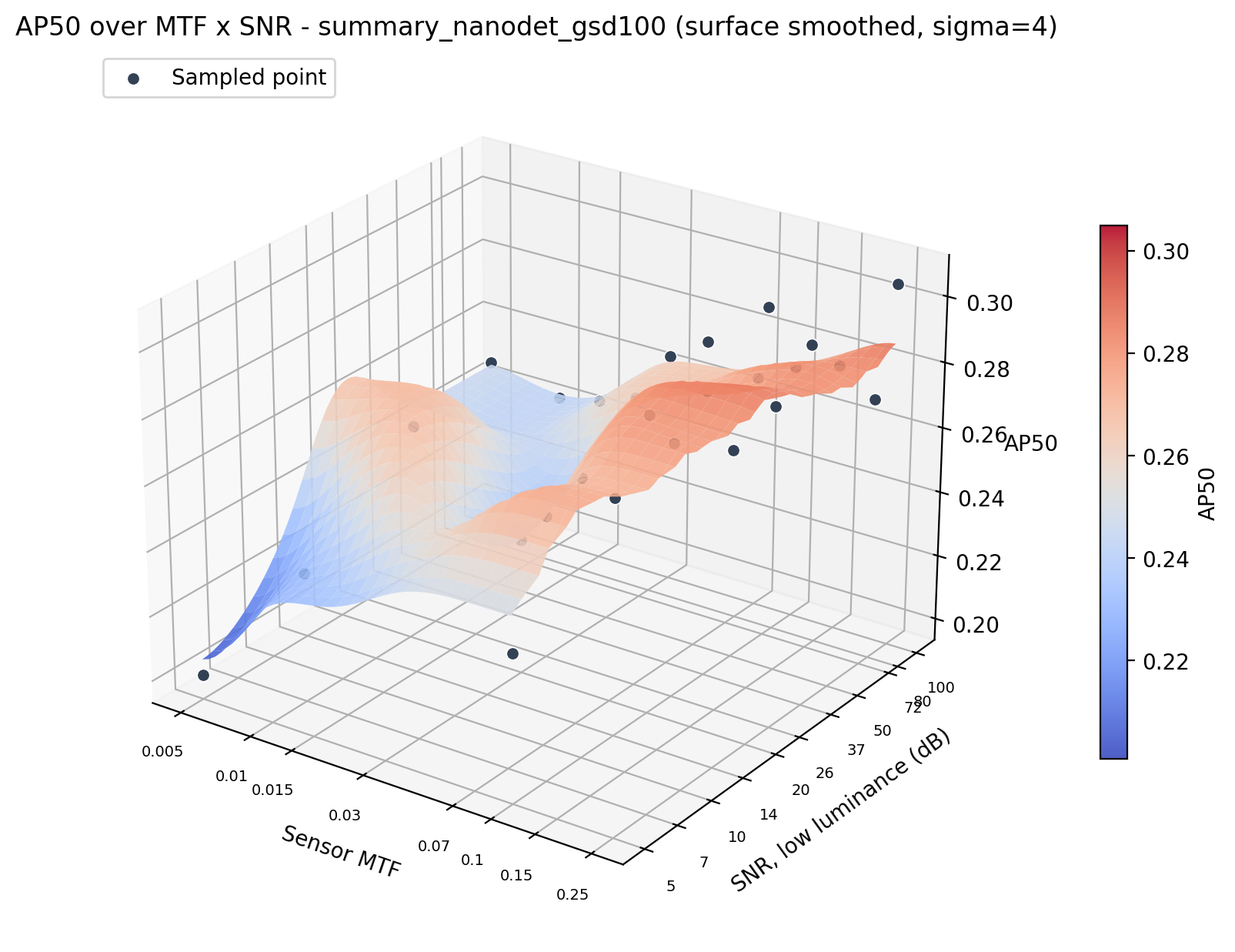}
        \caption{NanoDet AP$_{50}$ over the sampled parameter space.}
        \label{fig:nanodet_gsd100_surface}
    \end{subfigure}

    \begin{subfigure}[t]{0.49\textwidth}
        \centering
        \includegraphics[height=0.19\textheight,keepaspectratio]{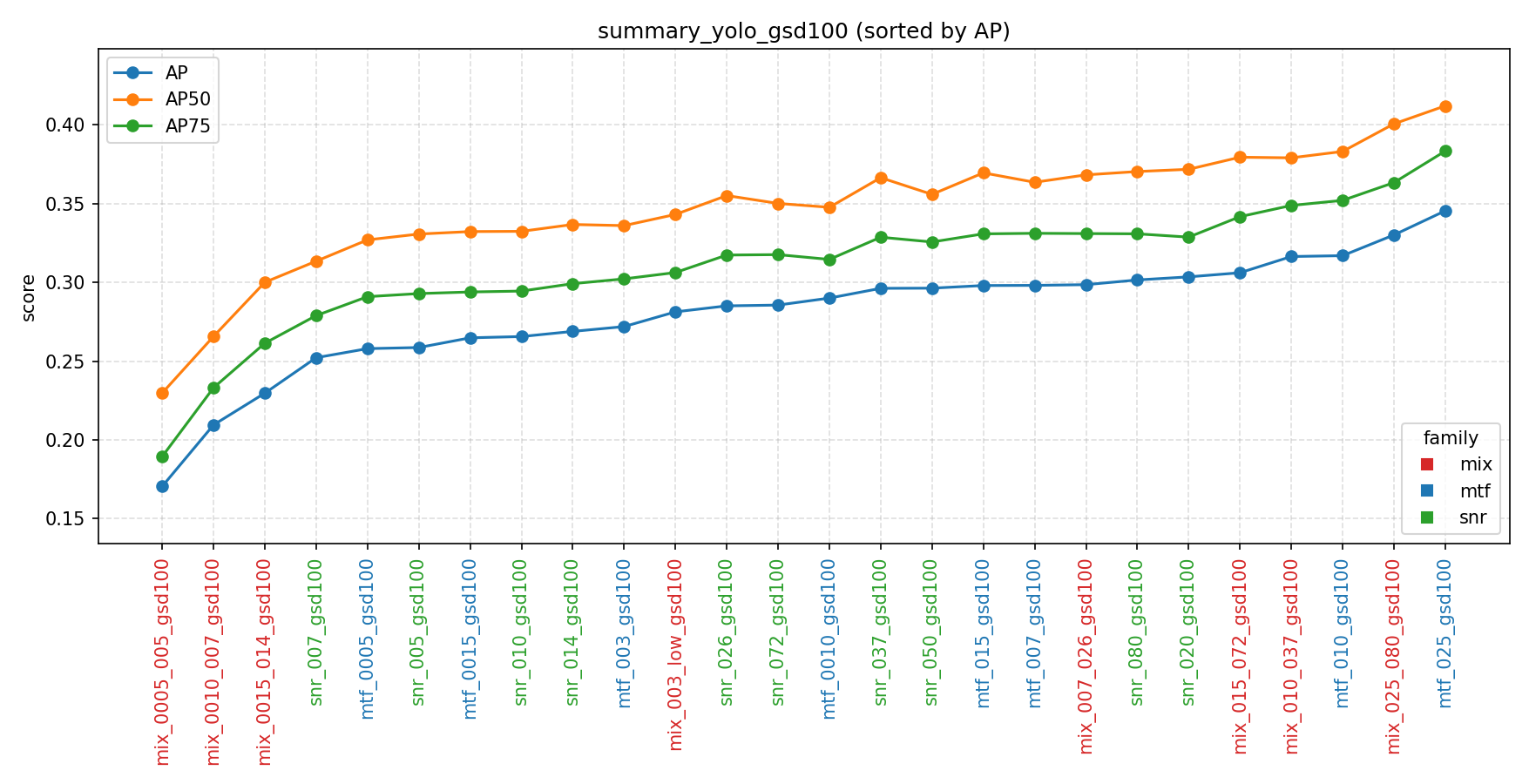}
        \caption{YOLOv5s configurations, ordered by AP.}
        \label{fig:yolo_gsd100_sorted}
    \end{subfigure}
    \hfill
    \begin{subfigure}[t]{0.49\textwidth}
        \centering
        \includegraphics[height=0.19\textheight,keepaspectratio]{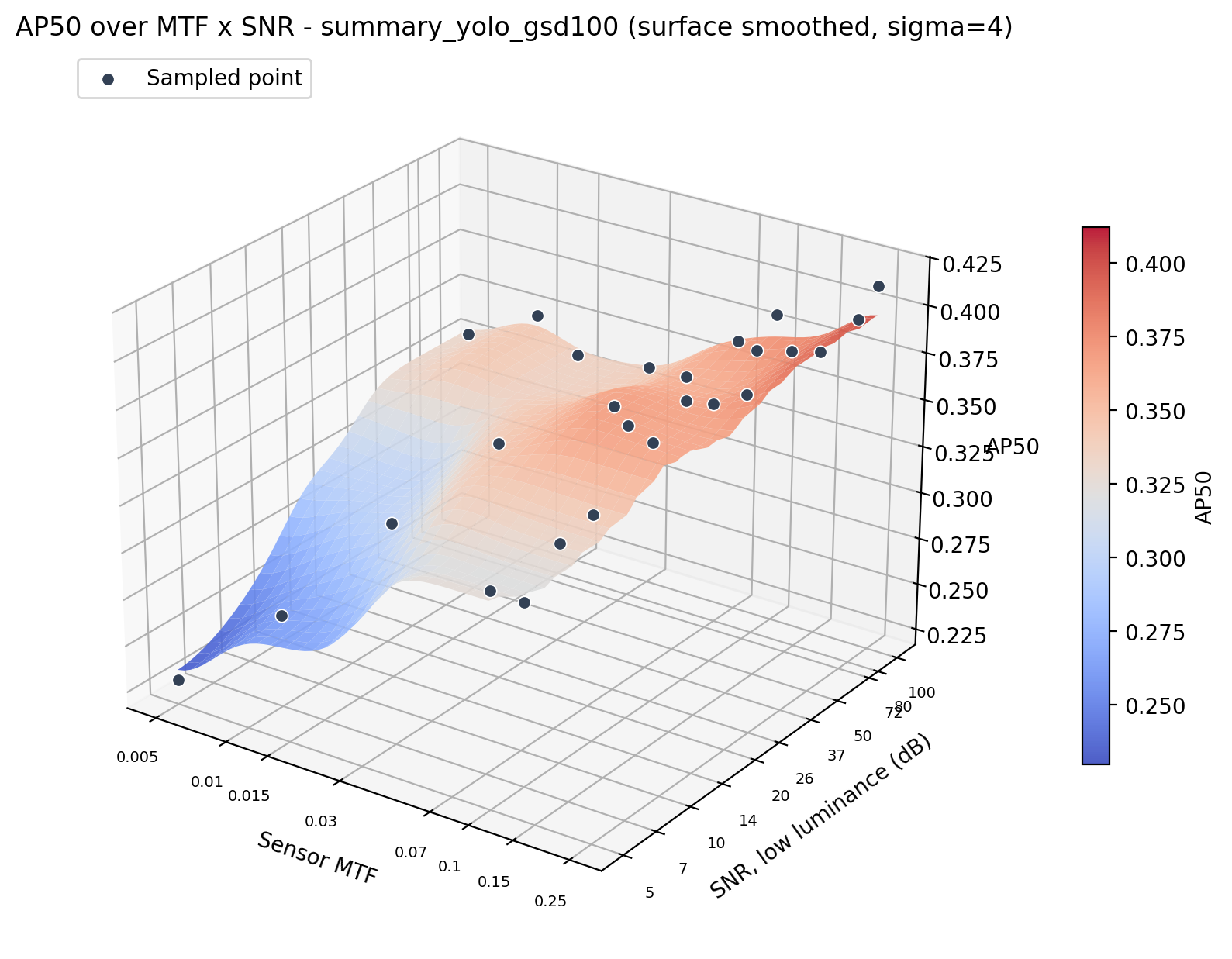}
        \caption{YOLOv5s AP$_{50}$ over the sampled parameter space.}
        \label{fig:yolo_gsd100_surface}
    \end{subfigure}

    \begin{subfigure}[t]{0.49\textwidth}
        \centering
        \includegraphics[height=0.19\textheight,keepaspectratio]{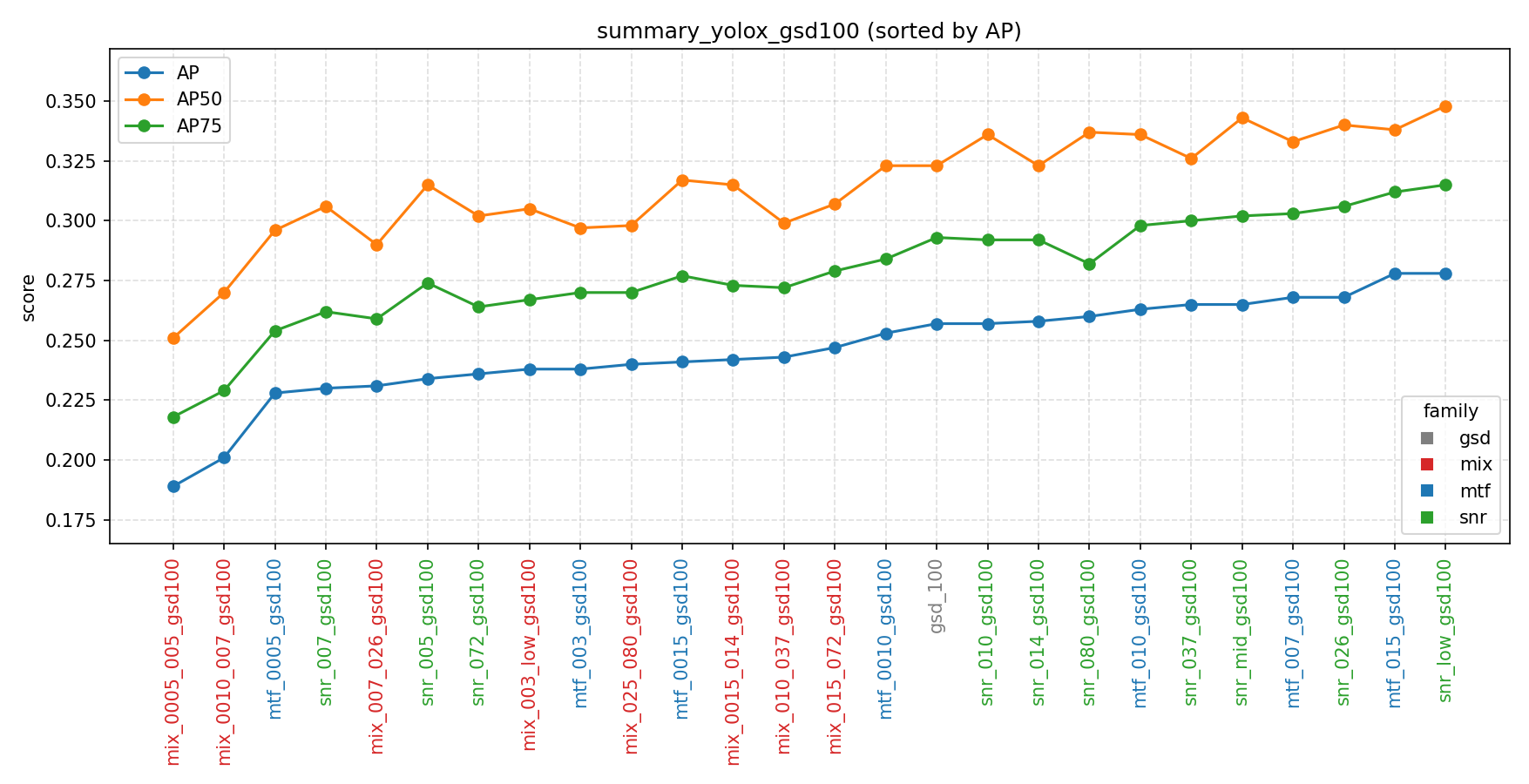}
        \caption{YOLOX-S configurations, ordered by AP.}
        \label{fig:yolox_gsd100_sorted}
    \end{subfigure}
    \hfill
    \begin{subfigure}[t]{0.49\textwidth}
        \centering
        \includegraphics[height=0.19\textheight,keepaspectratio]{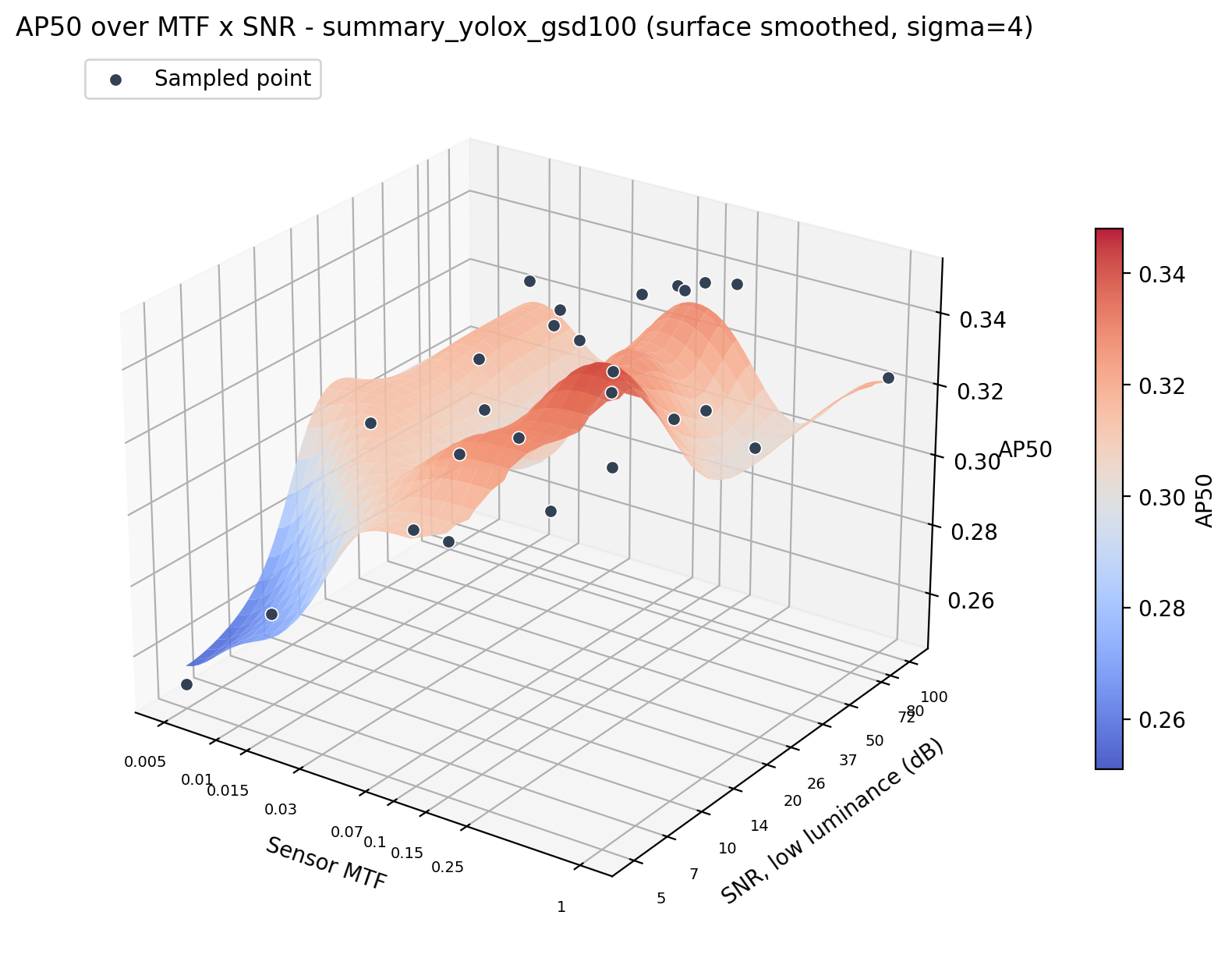}
        \caption{YOLOX-S AP$_{50}$ over the sampled parameter space.}
        \label{fig:yolox_gsd100_surface}
    \end{subfigure}

    \caption{Detection performance at GSD~=~100~cm. Left panels show the tested configurations and right panels show interpolated AP$_{50}$ surfaces for visualization.}
    \label{fig:results_gsd100}
\end{figure*}

\begin{figure*}[t]
    \centering

    \begin{subfigure}[t]{0.49\textwidth}
        \centering
        \includegraphics[height=0.19\textheight,keepaspectratio]{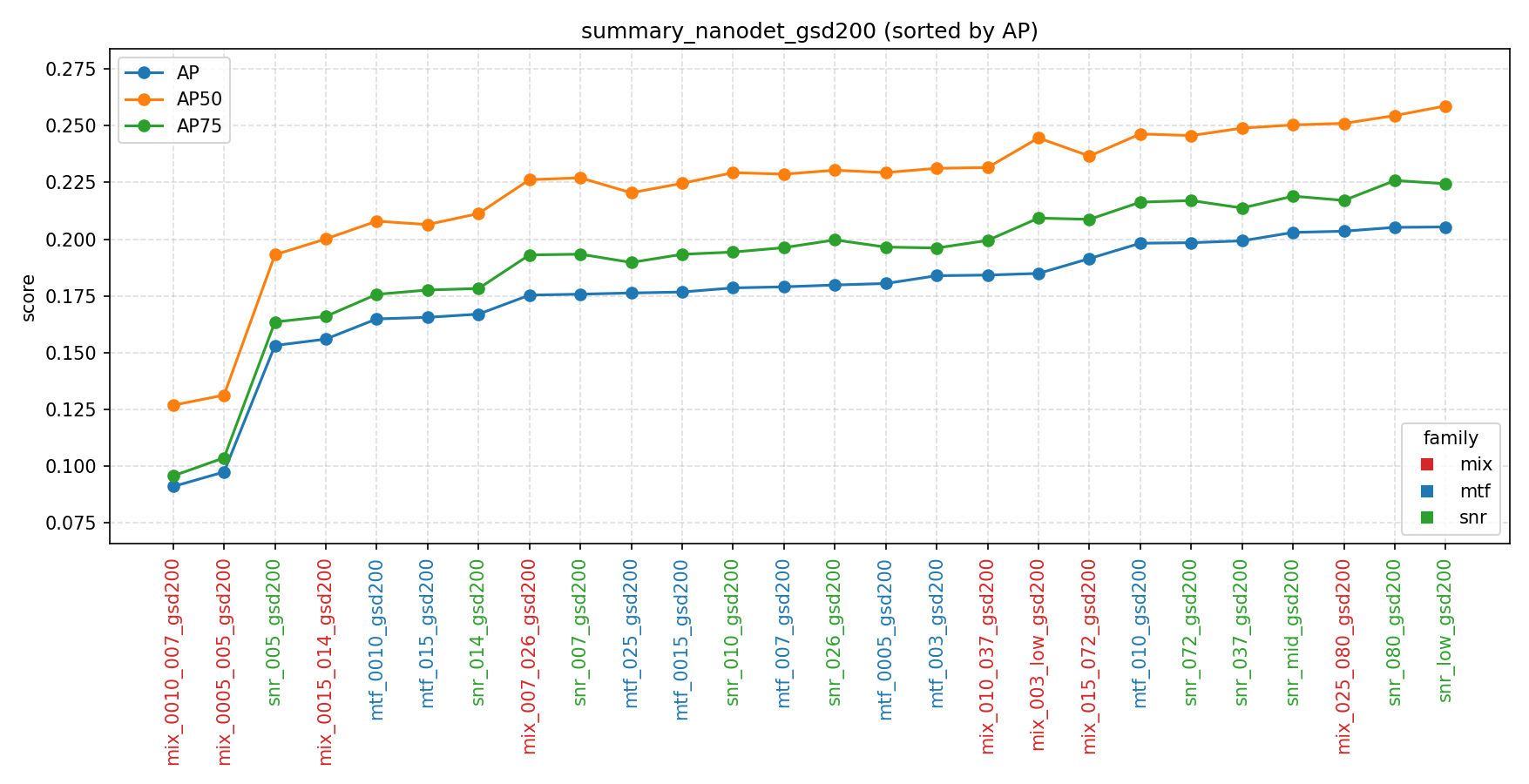}
        \caption{NanoDet configurations, ordered by AP.}
        \label{fig:nanodet_gsd200_sorted}
    \end{subfigure}
    \hfill
    \begin{subfigure}[t]{0.49\textwidth}
        \centering
        \includegraphics[height=0.19\textheight,keepaspectratio]{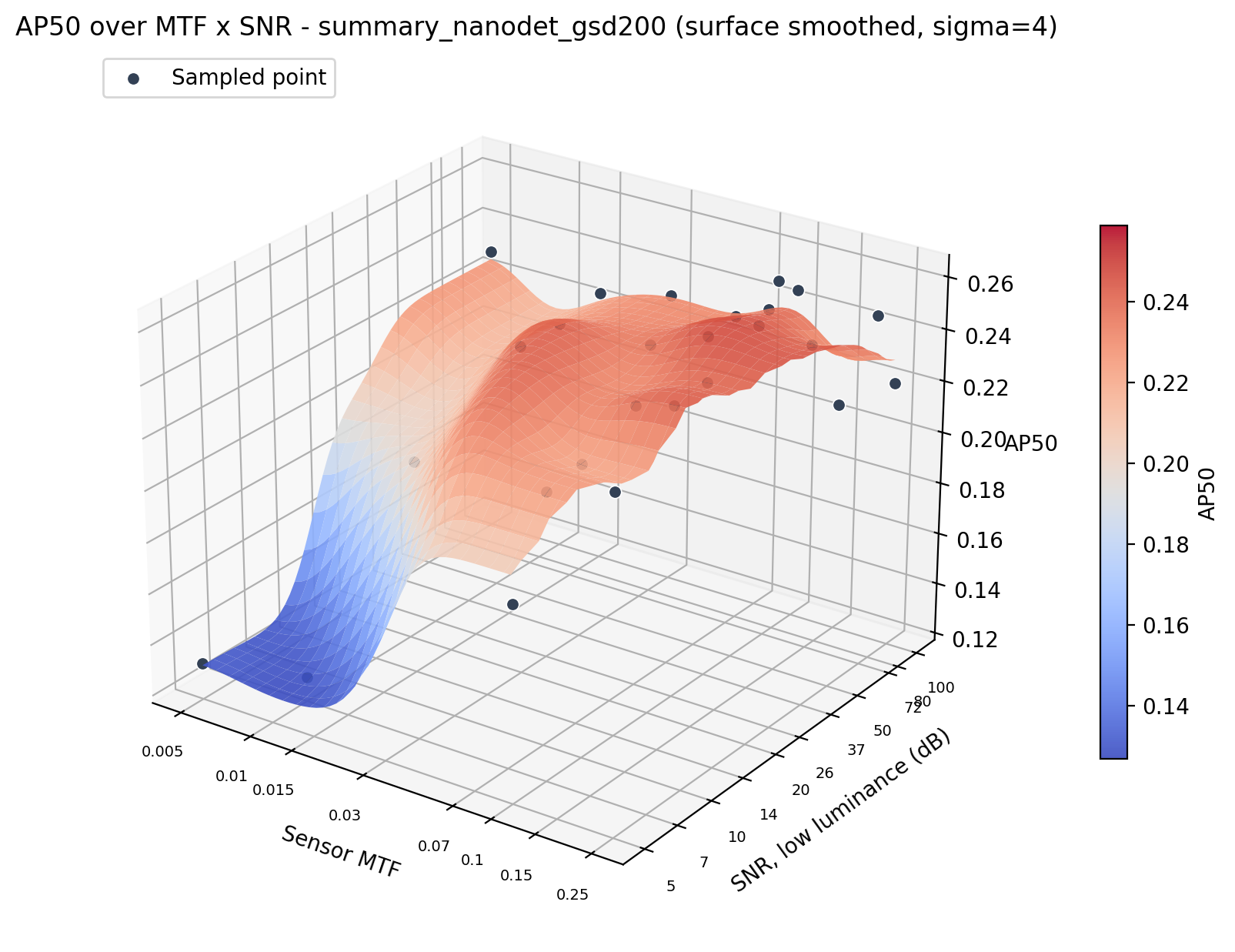}
        \caption{NanoDet AP$_{50}$ over the sampled parameter space.}
        \label{fig:nanodet_gsd200_surface}
    \end{subfigure}

    \begin{subfigure}[t]{0.49\textwidth}
        \centering
        \includegraphics[height=0.19\textheight,keepaspectratio]{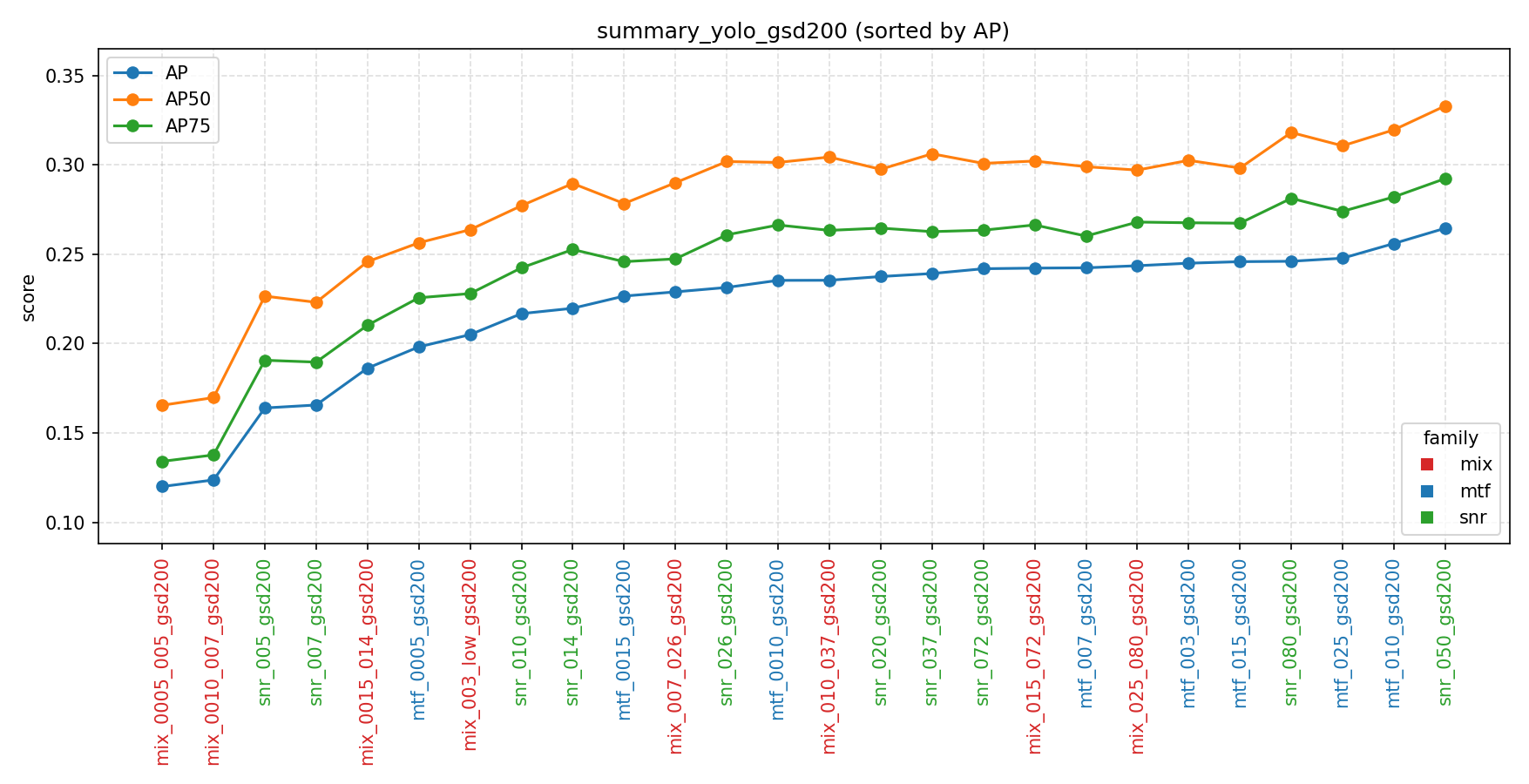}
        \caption{YOLOv5s configurations, ordered by AP.}
        \label{fig:yolo_gsd200_sorted}
    \end{subfigure}
    \hfill
    \begin{subfigure}[t]{0.49\textwidth}
        \centering
        \includegraphics[height=0.19\textheight,keepaspectratio]{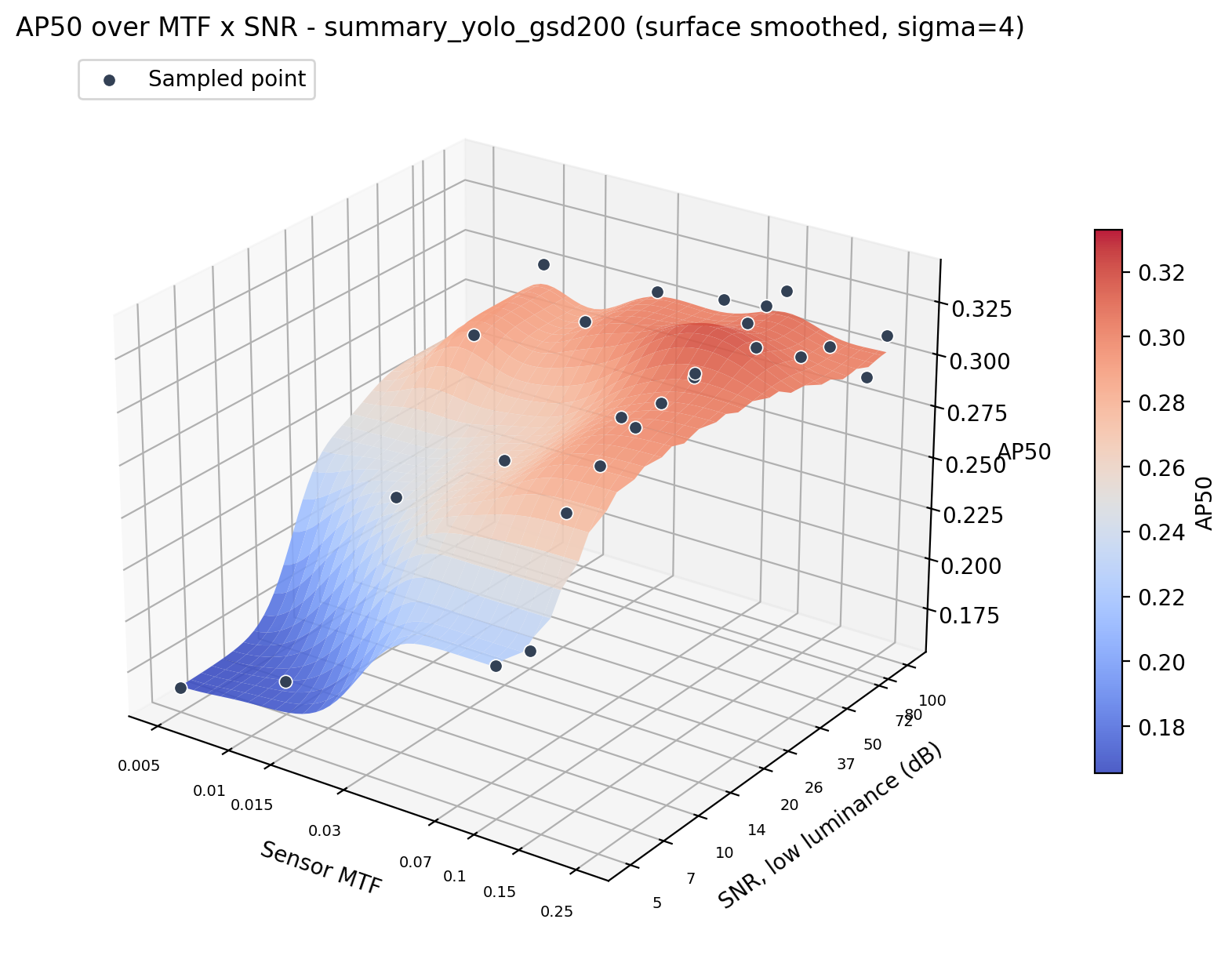}
        \caption{YOLOv5s AP$_{50}$ over the sampled parameter space.}
        \label{fig:yolo_gsd200_surface}
    \end{subfigure}

    \begin{subfigure}[t]{0.49\textwidth}
        \centering
        \includegraphics[height=0.19\textheight,keepaspectratio]{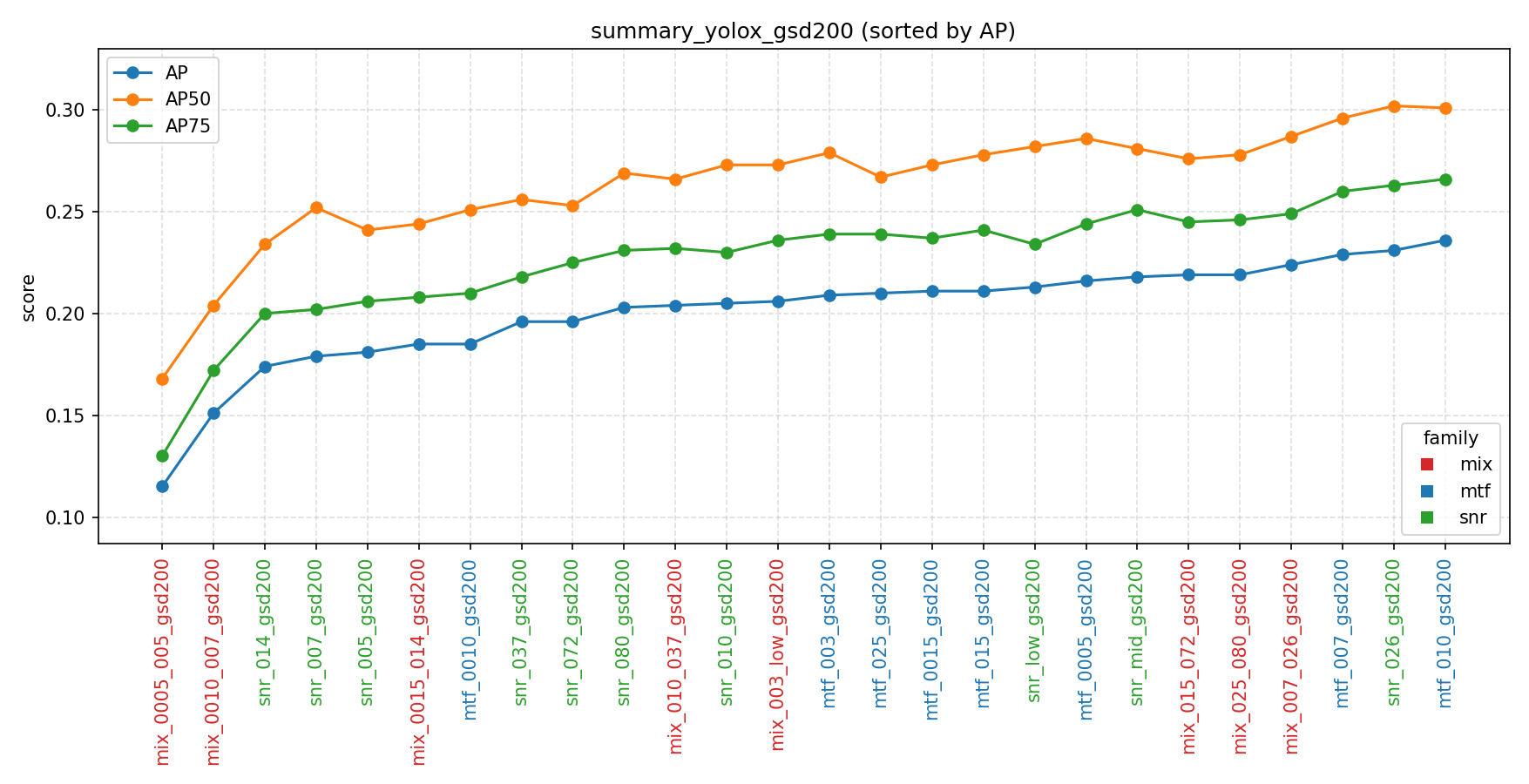}
        \caption{YOLOX-S configurations, ordered by AP.}
        \label{fig:yolox_gsd200_sorted}
    \end{subfigure}
    \hfill
    \begin{subfigure}[t]{0.49\textwidth}
        \centering
        \includegraphics[height=0.19\textheight,keepaspectratio]{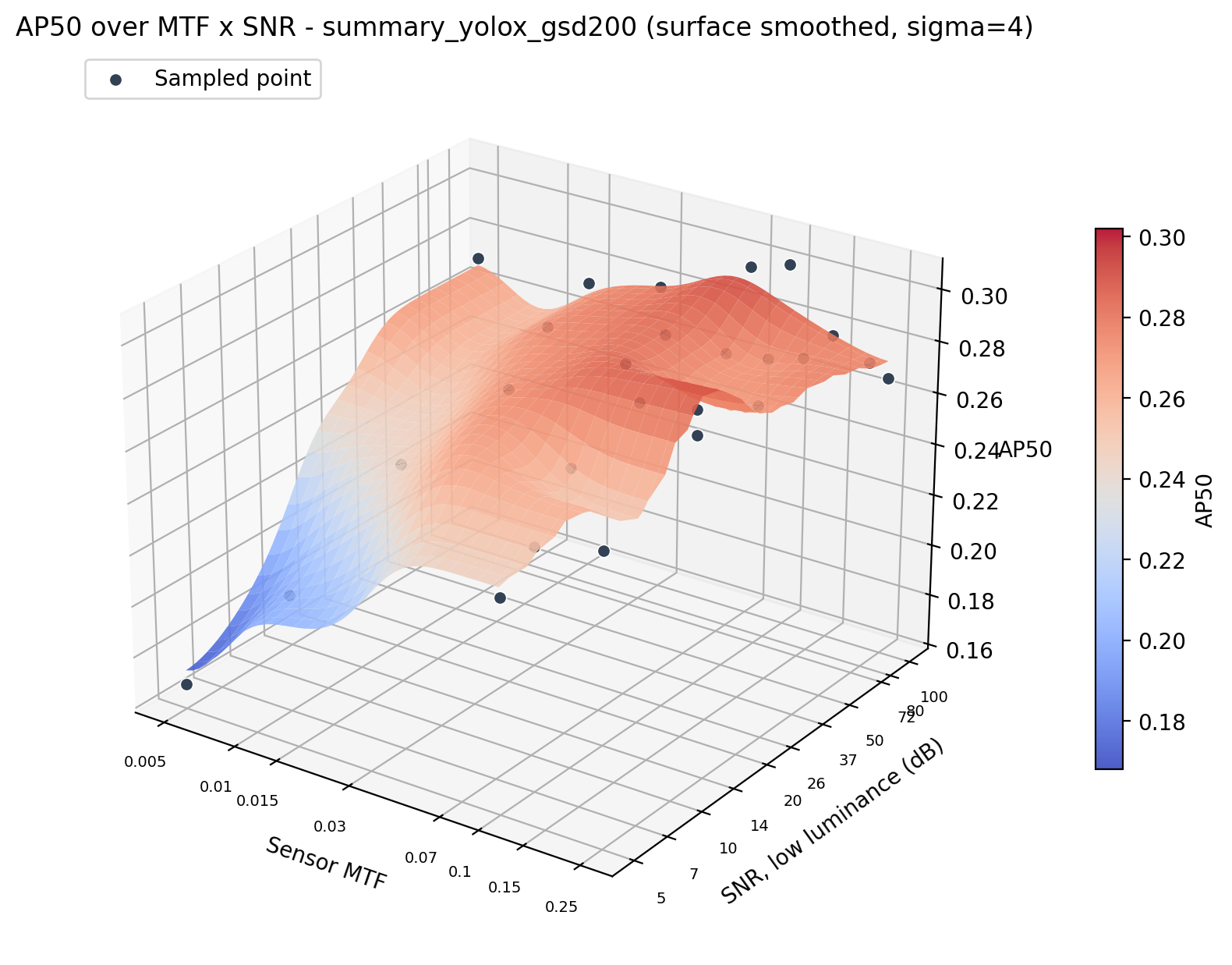}
        \caption{YOLOX-S AP$_{50}$ over the sampled parameter space.}
        \label{fig:yolox_gsd200_surface}
    \end{subfigure}

    \caption{Detection performance at GSD~=~200~cm. Left panels show the tested configurations and right panels show interpolated AP$_{50}$ surfaces for visualization.}
    \label{fig:results_gsd200}
\end{figure*}

\tabResults

\section{Discussion}

The experiments confirm that sensitivity to simulated raw-like image degradation depends on both the physical degradation mechanism and detector architecture. GSD produces the clearest overall effect. YOLOv5s and YOLOX-S perform better at 100~cm for every matched controlled configuration, with mean AP$_{50}$ differences of 6.7 and 5.0 points. NanoDet shows the same tendency in 22 of 23 matched cases. Spatial sampling is therefore the most consistent image-quality factor observed in this study.

At fixed GSD, the relative influence of MTF and SNR is less straightforward. At 100~cm, YOLOv5s and NanoDet show larger endpoint losses under severe MTF degradation than under the tested SNR degradation. Strong MTF degradation removes spatial frequencies and reduces edge and shape contrast that can be useful for object localisation and discrimination \cite{pleiades_restoration,raw_detection_edhpc}. At 200~cm, however, YOLOv5s becomes strongly sensitive to SNR, while NanoDet and YOLOX-S show more non-monotonic behaviour. The results therefore do not support a universal rule in which either MTF or SNR is always the dominant parameter.

The most consistent low-quality regime is the simultaneous presence of severe blur and noise. For every model and both GSDs, the strongest combined degradation gives lower AP$_{50}$ than the corresponding isolated MTF or SNR endpoint. This is relevant for raw and minimally processed satellite imagery, where several sensor-related limitations can occur simultaneously \cite{raw_detection_edhpc,pyraws_thraws}. Nevertheless, the mixed-sweep samples only eight paired operating points, so interpret the interpolated surfaces as visual aids rather than measurements of a complete MTF and SNR response surface.

Another important observation is the non-monotonic behaviour at moderate-to-high image quality. Several neighbouring operating points differ by only a few AP points and occasionally reverse the expected ordering. These variations may reflect a performance plateau and stochastic training variability. This plateau is visible in most of the 3D response surfaces in Figs.~\ref{fig:results_gsd100} and~\ref{fig:results_gsd200}, extending from the high-image-quality corner toward the central region of the explored parameter space. The main takeaway is that moderate image degradation does not necessarily translate into a substantial loss in detection performance. This suggests that some optical requirements could potentially be relaxed while maintaining comparable task-level performance. Because we evaluated the detectors across a broad range of operating points, we can also identify the image-quality region where performance starts to degrade substantially and use this threshold to inform sensor-design requirements.

From a system-design perspective, these results can help determine whether image restoration is required before onboard inference. Comparing detection performance on restored or ground-processed imagery with that obtained from raw or minimally processed imagery estimates the performance loss from operating directly on degraded sensor data.  If task-level accuracy remains sufficient over the expected sensor operating region, direct inference on minimally processed imagery may avoid the computational cost and latency of restoration \cite{pyraws_thraws,venus_dataset_full_author}. If performance is insufficient under expected degradation conditions, lightweight restoration or degradation-aware detection architectures may be more appropriate. Our previous work on ConvBEERS \cite{convbeers} investigates lightweight restoration as a preprocessing stage, while TriCCOT \cite{triccot} explores a degradation-aware detection architecture. The most appropriate solution therefore depends on the mission requirements, acceptable detection performance, and expected sensor operating conditions.

\section{Conclusion}

This work evaluates the impact of GSD, MTF, and signal-dependent radiometric noise on ship-detection performance in simulated raw-like optical imagery using YOLOv5s, NanoDet, and YOLOX-S. Among the investigated parameters, GSD produces the most consistent effect, with performance generally decreasing at 200~cm compared with 100~cm. In contrast, the effects of MTF and SNR are more dependent on spatial resolution, with several experiments exhibiting non-monotonic behaviour.

Overall, small-to-moderate degradations do not cause a sudden breakdown in detection performance, indicating that the evaluated detectors are robust within this degradation range. The most consistent failure regime across the three detectors occurs when severe blur and noise are combined. Under these conditions, AP$_{50}$ decreases more substantially than for the corresponding isolated degradations, highlighting the importance of considering interactions between image-quality factors rather than evaluating each parameter independently. More generally, the observed non-linear and model-dependent responses indicate that sensor requirements should not be derived from an assumed monotonic relationship between individual image-quality parameters and detection performance. Instead, detector performance should be characterised across the expected sensor operating domain. Such task-oriented evaluation can support early trade-offs between sensor design, onboard processing, and AI performance, and help determine whether direct inference on raw-like imagery, onboard restoration, or degradation-aware detection is most appropriate for a given mission.

A main limitation of this study is the absence of a statistical analysis of performance variability. Each configuration was evaluated using a single training run, making it difficult to distinguish systematic effects from variability associated with model training and potentially contributing to observed fluctuations and outliers. Future work should therefore include repeated experiments with multiple random seeds and report measures of variability or confidence intervals. This would provide a more robust assessment of the significance of small performance differences and help determine which trends can reliably be attributed to image degradation.

\bibliographystyle{abbrv}
\bibliography{biblio.bib}

@INPROCEEDINGS{versal_spaice_project,
  author={Garcés-Socarrás, Luis M. and Cuiman, Raudel and Ortiz, Flor and Vásquez-Peralvo, Juan A. and González-Rios, Jorge L. and Chehaitly, Mouhamad and Kazanskii, Arkadii and Malmir, Sahar and Nik, Amirhossein and Thoemel, Jan and Kumar, Sumit and Kuhfuss, Marcele and Varadajulu, Swetha and Lagunas, Eva and Duncan, Juan C. M. and Querol, Jorge and Chatzinotas, Symeon},
  booktitle={2025 European Data Handling \& Data Processing Conference (EDHPC)}, 
  title={Onboard Machine Learning for Satellite Edge Computing: The SPAICE Project Use Case}, 
  year={2025},
  volume={},
  number={},
  pages={1-8},

  doi={},
  }

@INPROCEEDINGS{versal_ship_detection_yolo,
  author={Ibrahim, Younis and Chen, Li and Haonan, Tian},
  booktitle={2022 14th International Conference on Computational Intelligence and Communication Networks (CICN)}, 
  title={Deep Learning-based Ship Detection on FPGAs}, 
  year={2022},
  volume={},
  number={},
  pages={454-459},
  doi={10.1109/CICN56167.2022.10008312}}

@ARTICLE{remote_sensing_review,
  author={Zhang, Bing and Wu, Yuanfeng and Zhao, Boya and Chanussot, Jocelyn and Hong, Danfeng and Yao, Jing and Gao, Lianru},
  journal={IEEE Journal of Selected Topics in Applied Earth Observations and Remote Sensing}, 
  title={Progress and Challenges in Intelligent Remote Sensing Satellite Systems}, 
  year={2022},
  volume={15},
  number={},
  pages={1814-1822},
  doi={10.1109/JSTARS.2022.3148139}}

@ARTICLE{cloud_detection_onboard,
  author={Aybar, Cesar and Mateo-García, Gonzalo and Acciarini, Giacomo and Růžička, Vít and Meoni, Gabriele and Longépé, Nicolas and Gómez-Chova, Luis},
  journal={IEEE Journal of Selected Topics in Applied Earth Observations and Remote Sensing}, 
  title={Onboard Cloud Detection and Atmospheric Correction With Efficient Deep Learning Models}, 
  year={2024},
  volume={17},
  number={},
  pages={19518-19529},
  doi={10.1109/JSTARS.2024.3480520}}

@article{opssat_meoni,
  title={The OPS-SAT case: A data-centric competition for onboard satellite image classification},
  author={Meoni, Gabriele and M{\"a}rtens, Marcus and Derksen, Dawa and See, Kenneth and Lightheart, Toby and S{\'e}cher, Anthony and Martin, Arnaud and Rijlaarsdam, David and Fanizza, Vincenzo and Izzo, Dario},
  journal={Astrodynamics},
  volume={8},
  number={4},
  pages={507--528},
  year={2024},
  publisher={Springer}
}

@ARTICLE{irma_imagini,

  author={Goudemant, Thomas and Francesconi, Benjamin and Aubrun, Michelle and Kervennic, Erwann and Grenet, Ingrid and Bobichon, Yves and Bellizzi, Marjorie},
  journal={IEEE Journal of Selected Topics in Applied Earth Observations and Remote Sensing}, 
  title={Onboard Anomaly Detection for Marine Environmental Protection}, 
  year={2024},
  volume={17},
  number={},
  pages={7918-7931},
  doi={10.1109/JSTARS.2024.3382394}}

@inproceedings{CIAR,
  TITLE = {{D{\'e}tection de navires embarquable {\`a} bord de satellites}},
  AUTHOR = {Goudemant, Thomas and Francesconi, Benjamin and Farhat, Houssem and Daniel, Lionel and Thiery, Olivier and Kervennic, Erwann and Girard, Adrien and Mzoughi, Seif},
  URL = {https://hal.science/hal-03881738},
  BOOKTITLE = {{Actes de la 4{\`e}me Conference on Artificial Intelligence for Defense (CAID 2022)}},
  ADDRESS = {Rennes, France},
  ORGANIZATION = {{DGA Ma{\^i}trise de l'Information}},
  SERIES = {Actes de la 4{\`e}me Conference on Artificial Intelligence for Defense (CAID 2022)},
  YEAR = {2022},
  MONTH = Nov,
  HAL_ID = {hal-03881738},
  HAL_VERSION = {v1},
}

@INPROCEEDINGS{raw_detection_edhpc,
  author={Dorise, Adrien and Bellizzi, Marjorie and Girard, Adrien and Francesconi, Benjamin and May, Stéphane},
  booktitle={2025 European Data Handling and Data Processing Conference (EDHPC)}, 
  title={Explaining raw data complexity to improve satellite onboard processing}, 
  year={2025},
  volume={},
  number={},
  pages={1-8},
  doi={}}

@INPROCEEDINGS{pyraws_maritime1,
  author={Del Prete, Roberto and Meoni, Gabriele and Salvoldi, Manuel and Barretta, Domenico and Graziano, Maria Daniela and Longépé, Nicolas and Renga, Alfredo},
  booktitle={IGARSS 2024 - 2024 IEEE International Geoscience and Remote Sensing Symposium}, 
  title={Enhanced Maritime Monitoring Via Onboard Processing Of Raw Multi-Spectral Imagery by Deep Learning}, 
  year={2024},
  volume={},
  number={},
  pages={1713-1717},
  doi={10.1109/IGARSS53475.2024.10641068}}

@article{pyraws_thraws,
   title={Unlocking the Use of Raw Multispectral Earth Observation Imagery for Onboard Artificial Intelligence},
   volume={17},
   ISSN={2151-1535},
   url={http://dx.doi.org/10.1109/JSTARS.2024.3418891},
   DOI={10.1109/jstars.2024.3418891},
   journal={IEEE Journal of Selected Topics in Applied Earth Observations and Remote Sensing},
   publisher={Institute of Electrical and Electronics Engineers (IEEE)},
   author={Meoni, Gabriele and Prete, Roberto Del and Serva, Federico and De Beusscher, Alix and Colin, Olivier and Longépé, Nicolas},
   year={2024},
   pages={12521–12537} }

@INPROCEEDINGS{jetson_space,
  author={Van Der Smissen, Sebbe},
  booktitle={2025 European Data Handling and Data Processing Conference (EDHPC)}, 
  title={Towards Efficient On-Board AI: Performance Benchmarks of the Jetson Orin NX for Space Applications}, 
  year={2025},
  volume={},
  number={},
  pages={1-7},
  doi={}}

@misc{yolov5,
  author = {Ultralytics},
  title = {{YOLOv5}: {A} state-of-the-art real-time object detection system},
  year = {2021},
  howpublished = {\url{https://docs.ultralytics.com}},
  note = {Accessed: 12/12/2025}
}

@misc{yolox,
      title={YOLOX: Exceeding YOLO Series in 2021}, 
      author={Zheng Ge and Songtao Liu and Feng Wang and Zeming Li and Jian Sun},
      year={2021},
      eprint={2107.08430},
      archivePrefix={arXiv},
      primaryClass={cs.CV},
      url={https://arxiv.org/abs/2107.08430}, 
}

@article{pleiades_restoration,
AUTHOR = {Latry, C. and Fourest, S. and Thiebaut, C.},
TITLE = {RESTORATION TECHNIQUE FOR PLEIADES-HR PANCHROMATIC IMAGES},
JOURNAL = {The International Archives of the Photogrammetry, Remote Sensing and Spatial Information Sciences},
VOLUME = {XXXIX-B1},
YEAR = {2012},
PAGES = {555--560},
URL = {https://isprs-archives.copernicus.org/articles/XXXIX-B1/555/2012/},
DOI = {10.5194/isprsarchives-XXXIX-B1-555-2012}
}

@ARTICLE{venus_dataset_full_author,
  author={Del Prete, Roberto and Salvoldi, Manuel and Barretta, Domenico and Longépé, Nicolas and Meoni, Gabriele and Karnieli, Arnon and Graziano, Maria Daniela and Renga, Alfredo},
  journal={IEEE Journal of Selected Topics in Applied Earth Observations and Remote Sensing}, 
  title={Enhancing Maritime Situational Awareness Through End-to-End Onboard Raw Data Analysis}, 
  year={2025},
  volume={18},
  number={},
  pages={16997-17018},
  doi={10.1109/JSTARS.2025.3584999}}

@misc{nanodet,
    title={NanoDet-Plus: Super fast and high accuracy lightweight anchor-free object detection model.},
    author={RangiLyu},
    howpublished = {\url{https://github.com/RangiLyu/nanodet}},
    year={2021}
}

@InProceedings{convbeers,
    author    = {Dorise, Adrien and Bellizzi, Marjorie and Hlimi, Omar},
    title     = {Rethinking Satellite Image Restoration for Onboard AI: A Lightweight Learning-Based Approach},
    booktitle = {Proceedings of the IEEE/CVF Conference on Computer Vision and Pattern Recognition (CVPR) Workshops},
    month     = {June},
    year      = {2026},
    pages     = {10183-10192}
}

@misc{triccot,
      title={TriCCOT: Tri-part Convolutional Conformal Transformer for Onboard Space Object Detection}, 
      author={Adrien Dorise and Marjorie Bellizzi and Julia Cohen and Stéphane May},
      year={2026},
      eprint={2609.08659},
      archivePrefix={arXiv},
      primaryClass={cs.CV},
      url={https://arxiv.org/abs/2609.08659}, 
}

@InProceedings{Goudemant_2026_CVPR,
    author    = {Goudemant, Thomas and Francesconi, Benjamin},
    title     = {Optimizing Latent Representations for Robust Building Damage Assessment Onboard Earth Observation Satellites},
    booktitle = {Proceedings of the IEEE/CVF Conference on Computer Vision and Pattern Recognition (CVPR) Workshops},
    month     = {June},
    year      = {2026},
    pages     = {10232-10240}
}

\end{document}